\documentclass[prc,english,aps,twocolumn,nofootinbib]{revtex4-1}
\usepackage[T1]{fontenc}
\usepackage[latin9]{inputenc}
\usepackage{amstext}
\usepackage{graphicx}
\usepackage{esint}
\usepackage{babel}
\begin{document}

\preprint{FTUAM-XX-2016}

\title{The Barcelona-Catania-Paris-Madrid functional with a realistic effective mass}

\author{M. Baldo}

\email{Marcello.Baldo@ct.infn.it}
\affiliation{Instituto Nazionale di Fisica Nucleare, Sezione di Catania, 
Via Santa Sofia 64, I-95123 Catania, Italy}

\author{L.M. Robledo}

\email{luis.robledo@uam.es}

\affiliation{Dep. F\'\i sica Te\'orica (M\'odulo 15), 
Universidad Aut\'onoma de Madrid,
E-28049 Madrid, Spain}

\author{P. Schuck}

\email{schuck@ipno.in2p3.fr}

\affiliation{Institut de Physique Nucl\'eaire, CNRS, UMR8608, F-91406 Orsay, France}
\affiliation{Universit\'e Paris-Sud, Orsay F-91505, France}
\affiliation{ Laboratoire de Physique et Mod\'elisation des Milieux Condens\'es, 
CNRS et Universit\'e Joseph Fourier, 25 Av. des Martyrs, BP 166, 
F-38042 Grenoble Cedex 9, France}

\author{X. Vi\~nas}

\email{xavier@ecm.ub.es}

\affiliation{Departament d'Estructura i Constituents de la Mat\`eria and Institut
de Ci\`encies del Cosmos, Facultat de F\'\i sica, Universitat de Barcelona,
Diagonal \emph{647}, 08028 Barcelona, Spain}


\begin{abstract}
The Barcelona-Catania-Paris-Madrid (BCPM) functional recently proposed
to describe nuclear structure properties of finite nuclei is generalized 
as to include a realistic effective mass. The resulting functional is as
good as the previous one in describing binding energies, radii, deformation
properties, etc and, in addition, the description of Giant Quadrupole Resonance
energies is greatly improved.
\end{abstract}

\maketitle


\section{Introduction}

In a recent paper \cite{BCPM}, we developed an energy density functional theory
for finite nuclei inspired in the Kohn Sham approach 
where the bulk part was fitted to the microscopic results of Baldo et al \cite{Bal10}.
They were obtained with the Brueckner Hartree-Fock (BHF) approach including
a three body force taken from Ref \cite{Akm98}. The interaction term of the equations of 
state (EOS) for symmetric nuclear and pure neutron
matters were represented by polynomials of even powers in the density
supplemented by a quadratic interpolation for asymmetric matter. In this way, a
very faithful representation of the microscopic energy per particle $E(\rho_p,
\rho_n)/A$ as a function of proton (p) and neutron (n) densities was obtained
for densities up to about three times saturation density ($\rho_0$).

In order to account for finite nuclei, a very simple Hartree type of term was
added with a single Gaussian as effective central two body force. Its strength
was fixed from the second order term of the polynomial fit and, thus, only one
free parameter, the range, was left for adjustment. A second adjustable
parameter was given by the strength $W_\mathrm{LS}$ of the spin-orbit force which was not
extracted from the microscopic calculation though, in principle, this might be
possible \cite{Sch76,Koh12,Fuj00}. A third parameter came from the fact that the microscopic
equilibrium value of the energy per particle had to be slightly renormalised by
about 10$^{-2}$ percent because in finite nuclei this value gets coupled with
the surface energy. In this case, all coefficients of the polynomials in $\rho_n, \rho_p$ have
then been changed by the same factor. With only those three adjustable
parameters, namely $r_0$, $W_\mathrm{LS}$ and $E/A$ of the infinite system, 
the rms deviation from experimental masses and charge radii were 1.58 MeV and 
0.027 fm, respectively \cite{BCPM}. 
An analogous procedure for constructing the functional
was followed in Ref \cite{Fay98}. A variant of this approach was adopted in Ref \cite{Cao06}, 
where a Skyrme force was derived from BHF calculations in nuclear matter, rather than directly the functional. A
peculiarity of the BCPM functional is that, like in Ref \cite{Fay98} but contrary to most of the Skyrme functionals,
its effective mass $m^*$ is equal to the bare one
$m =m^*$. The question of effective mass is a quite subtle one. In
principle there are two types of effective masses, the so-called $k$-mass and
the $\omega$-mass \cite{Jeu76}. The $k$-mass stems from the non locality and, thus, from the
momentum ($k$) dependence of the static Hartree-Fock type of mean field which, to
fix the ideas, may be derived from a Bruckner $G$-matrix \cite{BGM}. A typical value
of the effective $k$-mass is $m^* = 0.7 m$. On the other hand the so-called
$\omega$-mass is obtained in considering dynamic corrections to the single
particle self energy which lead to an energy ($\hbar \omega$) dependence. Most
of the time a coupling of the single particle motion to higher configurations or
to collective modes is considered \cite{Jeu76} and this compensates to a large
percentage the reduction of the $k$-mass with respect to the bare mass, so that
the combined effect is that the total effective mass becomes close to the bare
mass again. This effect, however, only holds for the states close to the Fermi
energy whereas in the calculation of the ground state energy all configurations
enter, so that a precise decision of whether one should take a reduced effective
mass or not is difficult to make unless one really undertakes a reliable
microscopic calculation for the $\omega$-mass and includes it in the
self-consistent mean field cycle. This, however, tremendously complicates the
whole approach. Anyway, it seems to be a fact that EDF's with or without reduced
masses are about equally successful and it must be concluded that apparently
the ambiguity of the effective mass can be very efficiently mocked up by a
renormalisation of finite size properties such as, e.g., the surface energy. On
the other hand, for excited states the story can be different. There, not only
the $\omega$-mass may be important in considering the propagation of a p-h
configuration but collective modes can also be exchanged between the particle
and the hole leading to a vertex correction. However, for monopole and isovector
giant resonances $\omega$-mass and vertex corrections cancel to a large extent,
see the review \cite{Ber.83} so that, again, for these modes
a bare mass can be taken in ph RPA calculations. The situation is different for
higher multipoles where this cancellation does not take place or only to a much
smaller extent.

It is our intention in this work to extend our BCPM-EDF to include an
effective density dependent mass which we extract again from the same $G$-matrix
calculations \cite{Bal04,Bal10} as it was done for the ground state energy. Also for proton
and neutron effective masses we will adjust a polynomial fit in the density. The
number of open parameters will stay the same is in the original BCPM-EDF. We
shall call the new EDF BCPM$^*$.

The paper is organized as follows, in Section \ref{sec:Methods} the methodology
used to introduce a non-constant effective mass in the BCPM functional is presented
along with details on the calculation of finite nuclei. In Section \ref{sec:Results}
we present the results of the fit of the new functional to the rms deviation 
of binding energies. With the new set of parameters we have performed some nuclear
structure calculations, like the evaluation of potential energy surfaces relevant to fission 
and the estimation of the excitation energies of Giant Monopole and Giant Quadrupole
resonances.


\section{Methods}
\label{sec:Methods}


The BCPM energy density functional derived in Ref.~\cite{BCPM} 
is inspired by the Kohn-Sham density functional theory \cite{KS}.
Although the original KS theory is local, it has been extended to
the non-local case, i.e. including effective mass and spin-orbit contributions 
(see Ref.~\cite{STV03} and references therein). It uses a simple polynomial of the
density $\rho$ and the isospin asymmetry parameter $\beta$ to fit the realistic
equation of state of symmetric and neutron matters obtained with a state of the
art microscopic calculation with realistic forces. We use the same polynomial 
for finite nuclei but this time in powers of the density of the finite nucleus
\begin{equation}
\rho (\vec{r}) = \sum_{ij} \phi^*_i (\vec{r}) \rho_{ij} \phi_j (\vec{r})
\end{equation}
Here the $\phi_i(\vec{r})$ are some basis wave functions (in our case, harmonic oscillator
wave functions) and $ \rho_{ij}$ is the density matrix.
To incorporate other effects not present or difficult to address in nuclear matter like the spin-orbit 
interaction or surface energy repulsion, additional terms discussed below are incorporated into
the functional. The kinetic energy is treated at the quantum mechanical level
by introducing the kinetic energy density
\begin{equation}\label{eq:quantumtau}
\tau (\vec{r}) = \sum_{ij} \vec{\nabla} \phi^*_i (\vec{r}) \rho_{ji} \vec{\nabla} \phi_j (\vec{r}).
\end{equation}
The total energy of a finite nucleus is then given by 
\begin{equation}
E = T + E_{int}^{\infty} + E_{int}^{FR} + E_{s.o} + E_C,
\label{eq:8}
\end{equation}
where $T$ is the kinetic energy, 
\begin{equation}
E_{int}^{\infty} = \int d\vec{r}\rho(\vec{r})\big[ P_s(\vec{r})(1 -\beta^2(\vec{r})) 
+ P_n(\vec{r})\beta^2(\vec{r})\big],
\end{equation}
is the bulk energy, given in terms of the polynomials $P_s(\rho)$ and $P_n(\rho)$ for symmetric
and neutron matter and the asymmetry density 
$\beta(\vec{r})=(\rho_n(\vec{r}) -\rho_p (\vec{r}))/\rho (\vec{r})$, 
$E_{int}^{FR}$ is a finite range surface term, $E_{s.o.}$ is the spin-
orbit energy taken from the Skyrme or Gogny forces and $E_C$ is the standard Coulomb repulsion
including the exchange energy in the Slater approximation. 
This energy is supplemented by a density-dependent zero range pairing interaction and some beyond
mean-field corrections (see again ref.~\cite{BCPM} for details). 

The inclusion of an effective mass in BCPM is carried out by means of
adding and subtracting an appropriate kinetic energy term to the original
kinetic energy density
\begin{equation}
\frac{\hbar^2}{2m} \tau \rightarrow \frac{\hbar^2}{2m^*} \tau  + 
\frac{\hbar^2}{2m}
\left[1-\left(\frac{m}{m^*}\right)\right] \tau^\infty.
\label{eq:kinem}
\end{equation}
In this expression $m^*$ is
the coordinate dependent effective mass, $\tau$ is the quantum kinetic
energy density of Eq. (\ref{eq:quantumtau}), and $\tau^\infty$ is the semiclassical
kinetic energy $\tau^\infty=\frac{3}{5} (3\pi^2)^{2/3} \rho^{5/3} (\vec{r})$.
This substitution guarantees that the kinetic energy at nuclear matter level
remains the same as before. For simplicity, no explicit
mention of isospin is made in the previous formulas, but different 
effective masses for protons and neutrons are considered.
With this redefinition of the kinetic energy the functional now reads
\begin{equation}
E = T^* + E_{int}^{\infty\,*} + E_{int}^{FR} + E_{s.o} + E_C
\label{eq:BCPMS},
\end{equation}
where
\begin{equation}
T^*= \sum_\tau \int d\vec{r} \frac{\hbar^2}{2m^*_\tau} \tau_\tau (\vec{r}),
\end{equation}
and $E_{int}^{\infty\,*}$ is obtained by adding
\begin{equation}
 \frac{\hbar^2}{2m} \sum_\tau \int d\vec{r} 
\left[1-\left(\frac{m}{m^*_\tau}\right)\right] \tau^\infty_\tau
\end{equation}
to $E_{int}^\infty$. The rationale behind this procedure is to preserve
the nuclear matter EoS of BCPM and, therefore, all the nuclear matter parameters
of this functional -- see \cite{BCPM} for a discussion.
  
Pairing correlations, required to describe open shell nuclei,
are introduced by means of a density-dependent zero-range force of 
the type suggested by Bertsch and Esbenssen \cite{ber91}. This force 
is widely used in nuclear structure calculations \cite{gar99,baldo08,baldo10}.
The pairing strength parameters are taken again from  \cite{gar99}. However in this 
work we choose the set of values corresponding to an efective mass different from the bare mass.
The two-body kinetic energy correction, which accounts for the lack of traslational 
invariance, is taken as in \cite{BCPM}. The final ingredient of the energy is the rotational
energy correction $\epsilon_\mathrm{rot} = \frac{\langle \vec{J}^2 \rangle}{2\mathcal{J}_Y}$
which is subtracted from the functional's energy. The rotational correction is defined in 
terms of the Yoccoz moment of inertia \cite{Ring.80} and computed using the HFB like intrinsic
wave function corresponding to the minimum of the HFB energy.
This procedure corresponds to the projection after variation (PAV) method applied to 
rotational symmetry restoration -- see 
\cite{Ring.80,RRG00,egido04} for a thorough explanation.
Note that the rotational energy
correction plays an important role in deformed nuclei and its inclusion is
relevant to describe masses along the whole periodic table.
In strongly deformed mid-shell heavy nuclei the rotational energy 
correction can reach values as large as 6 or 7 MeV. This correction, however, is 
almost negligible in magic or semi-magic nuclei, which are basically 
spherical. Due to the fact that the spherical-deformed transition is sharp,
the rotational correction goes from zero to some MeV at the transition point leading
to sharp variations in the binding energy plot -- see below.

The finite nuclei calculations have been carried out preserving axial symmetry and
using an adaptation of the computer code HFBaxial \cite{HFBaxial,rob11}. The quasiparticle
operators are expanded in a harmonic oscillator basis with varying number of oscillator
shells depending on mass number as to guarantee a weak dependence  of binding energies 
with the basis size -- see \cite{BCPM} for details. In \cite{BCPM} we have used a simple 
formula to extrapolate binding energies to the value corresponding to an infinite size HO basis.
As this procedure has proven to lead to some difficulties, we have preferred to increase the 
maximum basis size by adding two major shells, avoiding in this way the infinite basis extrapolation. 

\begin{figure}
\includegraphics[width=\columnwidth]{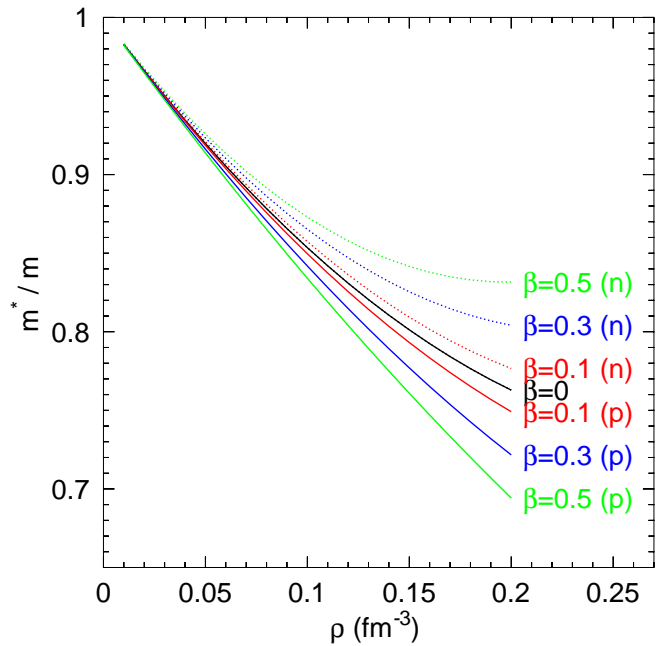}
\caption{Proton (p, full line) and neutron (n, dotted line) effective masses corresponding to the linear
fit of Eq. (\ref{eq:lin}) are plotted as a function
of the density for different values of the isospin asymmetry parameter $\beta$.\label{fig:m_star}}
\end{figure}
 
The effective masses $m^*_n$ and $m^*_p$ for neutrons
and protons are obtained in the
uniform system in terms of the  
neutron and proton single particle potentials $U_n$ 
and $U_p$, calculated within the Brueckner Hartree Fock (BHF) procedure.
At the Fermi momenta $k_{F q}$ one has
\begin{equation} 
 \frac{m_q}{m^*_q} \,=\, 1 \,+\, \frac{m_q}{\hbar^2 k_{F q}}\, \bigg( \frac{d U(k)}{d k}\, \bigg)_{k = k_{F q}}
\label{eq:effm}\end{equation}
\noindent
where $q = n,\, p$ and $m_q$ is the bare nucleon mass.
The effective mass is a function both of the total density $\rho$
and of the asymmetry $\beta \,=\, (\rho_n - \rho_p)/\rho$. \par
We used Eq. (\ref{eq:effm}) in a systematic calculation of the neutron
and proton effective masses for a set of total densities,
ranging from $0$ to $0.2$ fm$^{-3}$, and asymmetries
$\beta$ from $0$ to $1$. We found that the neutron and proton
effective masses can be fitted by a simple polynomial
expression
\begin{equation}
 \frac{m^*_n}{m_n} \,=\, a_0(\rho) \,-\, a_1(\rho) \beta \ \ \ \ \ ; \ \ \ \ \frac{m^*_p}{m_p}\,=\, a_0(\rho) \,+\, a_1(\rho) \beta,
\label{eq:lin}\end{equation}
\noindent
where
\begin{eqnarray}
\, &a_0 &=\, 1 - 1.744025 \rho + 2.792075 \rho^2 \nonumber \\
\, &a_1 &=\, 0.090795 \rho + 2.981724 \rho^2.
\label{eq:a01}\end{eqnarray}
\par 
This expression can be extended to negative values of $\beta$,
provided we simply interchange neutrons and protons, as it
must be.
Since the neutron and proton effective
masses have a symmetric splitting, see Eq. (\ref{eq:lin}) and Fig \ref{fig:m_star}, they 
have a continuous derivative at $\beta = 0$. The fit with 
the linear dependence of $\beta$ looks not enough for
the whole range up to $\beta = 1$ (pure neutron matter).
However a good fit can be obtained up to values of $\beta \approx 0.5$,
which are within the range of values appearing in stable
nuclei, with the exception of very light nuclei, for which
the BCPM functional is not expected to be applicable.
Therefore, we keep the fit of Eq. (\ref{eq:lin},\ref{eq:a01})
neglecting the deviations, which can appear at very large
asymmetries. These deviations could be taken into account
by introducing higher powers in $\beta$. However this would not be
relevant for finite nuclei, and we prefer to keep the simplicity of the
linear dependence. We are aware, however, that in situations of large asymmetry like
in Wigner-Seitz cells in neutron stars our simple linear dependence would not be enough. 
The present approach is in line with phenomenological
optical model analyses. In ref. \cite{Bao1} the neutron to proton
mass splitting is expressed as $(m^*_n - m^*_p)/m = (0.27 \pm 0.25) \beta$,
while more recently \cite{Bao2} a similar analysis gives $(0.41 \pm 0.15) \beta$.
From Eq. (\ref{eq:a01}) at saturation one finds $0.2 \beta$,
i.e. a splitting that agrees in the sign but it appears to be
on the small side. In any case one should appreciate 
the rough agreement between phenomenology and theory what is not obvious 
nor trivial. In finite nuclei the polynomial fit in Eq. (\ref{eq:lin}) is maintained but using
the finite nucleus density instead of the nuclear matter one.


\section{Results}
\label{sec:Results}


\begin{figure*}
\includegraphics[angle=-90,width=\textwidth]{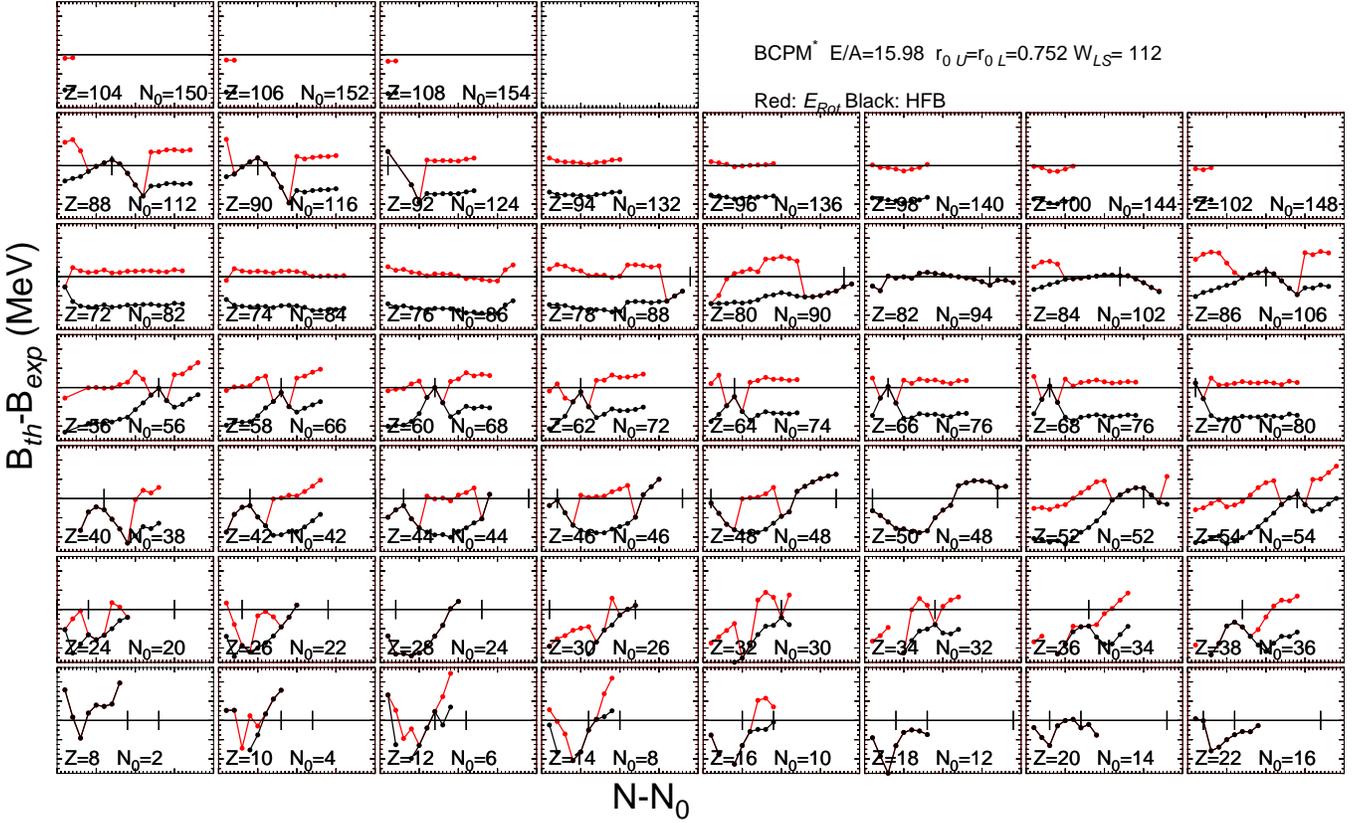}
\caption{The binding energy difference $\Delta B=B_\mathrm{th}-B_{\mathrm{exp}}$ 
(MeV) is plotted as a function of the shifted (by N$_{0}$(Z)) neutron number N-N$_{0}$(Z) 
for all the isotopic chains considered. The values of Z and the neutron number
shift N$_{0}$(Z) for each chain are given in the corresponding panel.
The ordinate $\Delta B$ axis ranges from -5.5 MeV to 5.5 MeV with long ticks every 2 MeV.
The N-N$_{0}$(Z) axis spans a range of 40 units with long ticks every 10 units
and short ones every 1 unit. In every panel, a horizontal 
line corresponding to $\Delta B=0$ has been plotted to 
guide the eye. Additional perpendicular lines signaling the position of 
magic neutron numbers have also been included.\label{fig:BE}}
\end{figure*}

In this section we discuss first the fitting of the free parameters of the BCPM$^*$
functional to minimize the root mean square (rms) binding energy difference with the
experimental data. The functional so obtained is then used to carry out
calculations to asset the merits of the functional regarding quadrupole deformation
properties and excitation energies of Giant Resonances.

\subsection{Binding energies and radii}

As a consequence of the introduction of the effective mass in finite nuclei, a
readjustment of the parameters of BCPM is required. The most affected is the spin-orbit
interaction strength, which is inversely proportional to the effective mass and, 
therefore, is going to have a value in BCPM$^*$ closer to the value of other functionals
like Gogny D1S or D1M with effective masses not equal to the bare one. The reason for such dependence
is the link between the spin-orbit strength and the magic numbers: $W_\mathrm{LS}$ has to
be large enough as to bring intruder orbital down to the lower major shell. Decreasing
the effective mass increases  the gap between major shells and, therefore, a larger $W_\mathrm{LS}$ 
value is required. The spin-orbit
strength along with the other two range parameters $r_{0L}$ and $r_{0U}$ (see \cite{BCPM} 
for more information) are readjusted as to fit the binding energies of even-even nuclei in 
a similar manner as in \cite{BCPM}. The minimum value of the rms for the binding energy difference using 
the AME 2012 experimental compilation including 620 even-even nuclei \cite{AW.12}
is $\sigma_\mathrm{E}= 1.68 \mathrm{MeV}$ which is
slightly higher than the original BCPM value of 1.58 MeV obtained with only 579 even-even nuclei
(1.61 MeV when the AME 2012 compilation is considered). The values of the fitted parameters
are $r_{0 U}=r_{0 L}=0.752 $ fm and $W_\mathrm{LS}=112$ MeV. We observe that as in the BCPM case, the
minimization of the binding energy favors equal values of the  $r_{0U}$ and $r_{0L}$ ranges. In Fig \ref{fig:BE} the 
binding energy difference is plotted as a function of neutron number for the
different values of Z  (see figure caption for an explanation of the
plot). In this plot we observe a nice reproduction of experimental data for heavy
nuclei away from magic or semi-magic numbers. Close to magic numbers we observe in many cases
a non-smooth behavior  which is due, as explained in \cite{BCPM},
to a deficiency on the way the rotational energy correction used in the binding energy
is computed: As mentioned in the previous section,
the rotational energy correction is obtained using the projection after variation
method where the correction is computed using the intrinsic wave function minimizing
the HFB energy and, therefore, it is zero for spherical intrinsic states. The correction suddenly
jumps by a couple of MeV when the spherical to deformed transition takes place and the jump
obviously reflects in the binding energy. This 
deficiency could be cured by computing the rotational energy correction in the variation 
after projection (VAP) scheme but this procedure, even in an approximate way, is much more 
costly to implement than the used PAV method. Work to find a convenient way to implement
the VAP is under way. For light nuclei the agreement with the experimental binding
energies deteriorates and the dependence with proton and neutron number is not well reproduced.

\begin{figure}
\centerline{\includegraphics[width=\columnwidth]{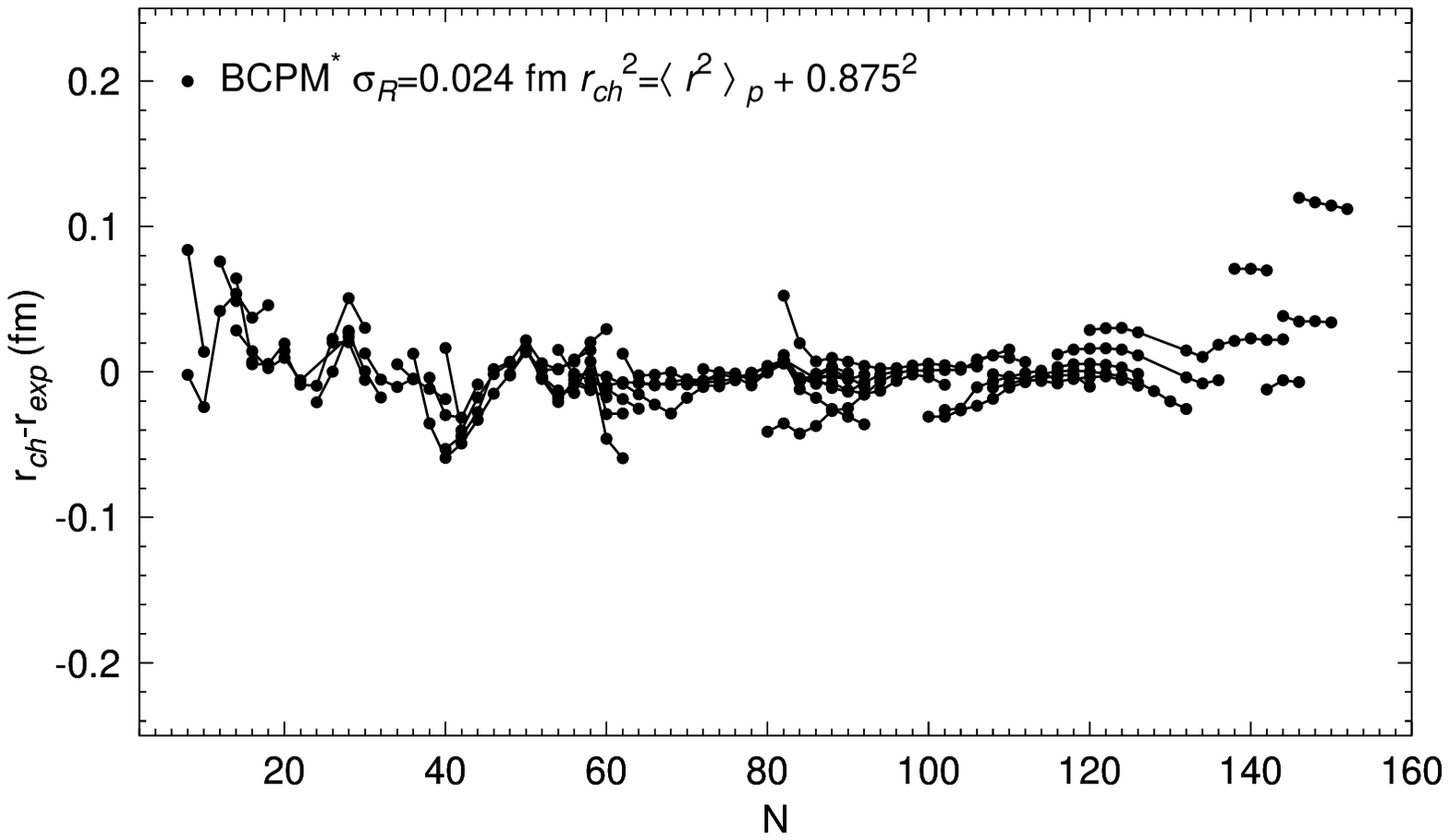}}
\caption{\label{fig:RA} The experiment-theory deviation $r_\mathrm{ch}-r_\mathrm{exp}$ 
for the 315 even-even nuclei with known experimental data \cite{angeli04} is plotted as a function
of mass number A.} 
\end{figure}

Concerning charge radii
we have also computed the root mean deviation $\sigma_R$ with respect to the 315 experimental
points in the recent compilation of Angeli et al \cite{angeli04}. The theoretical
radius is computed using the standard formula 
$r_\mathrm{ch} = \sqrt{\langle r^2 \rangle_\mathrm{HFB}+ 0.875^2}$. The value obtained 
for $\sigma_R$ using the BCPM$^*$ functional is
$\sigma_R = 0.024 \mathrm{fm}$, which is around 15\%  better than the 0.027 fm value obtained
with BCPM. In Fig \ref{fig:RA} we plot the difference between the theoretical and experimental value of the
charge radii as a function of the mass 
number A for the 315 even-even nuclei with experimentally known charge radii \cite{angeli04}. Overall,
we see a very good agreement with experimental data, except in some super-heavy and light nuclei. These
deficiencies were also observed in the BCPM results of Ref \cite{BCPM}.

As the nuclear matter EoS of BCPM has been preserved in BCPM$^*$, all its nuclear
matter parameters $K$, $J$, etc  remain exactly the same as with BCPM and we refer the reader
to Ref \cite{BCPM} for an extensive discussion of their values. In addition, the BCPM$^*$ values 
of the range parameters of the surface term have not changed substantially, with
respect to the ones of BCPM and, therefore, it is to be expected that 
the variance analysis of $\sigma_E$ with respect 
to the parameters $r_{0 L}$, $r_{0 U}$ and $W_\mathrm{LS}$ carried out in \cite{BCPM}
is going to yield similar conclusions for BCPM$^*$.

\subsection{Fission barrier heights}

\begin{figure}
\centerline{\includegraphics[width=\columnwidth]{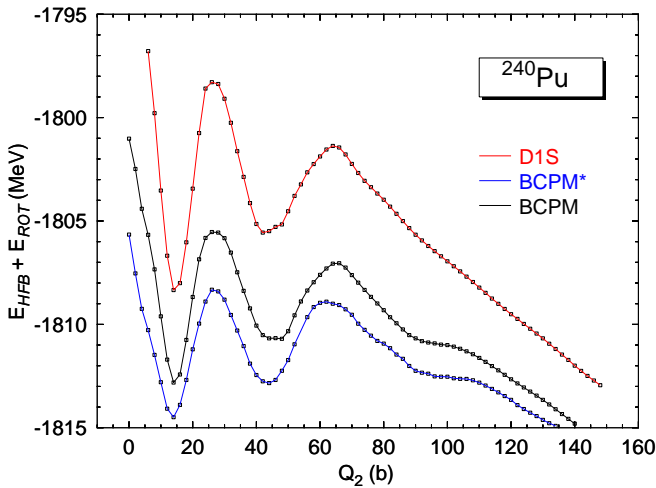}}
\caption{\label{fig:fission} Potential energy surface for fission, including the 
rotational energy correction, computed as a function of the quadrupole
moment (in b) for the Gogny D1S, BCPM and BCPM$^*$ functionals. }
\end{figure}

A fundamental aspect of any nuclear effective interaction is its ability to
produce reasonable deformation properties. The fission phenomenon,
which is described as the collective evolution of the nucleus from its ground 
state to scission using the quadrupole deformation parameter as driving coordinate,
is perhaps the best testing ground in this respect. From the perspective of
comparing with experimental data, there are well established values of the fission
barrier heights in a bunch of actinides and super-heavies that could be used. Those
values are extracted in a model dependent way from the behavior of the induced 
fission cross section as a function of the energy and are routinely used as 
benchmarks of theoretical fission models. In previous studies \cite{Giu.13,Giu14} we have shown that
the BCPM interaction produces quite reasonable results for fission observables.
Therefore, we have repeated some of the calculations to evaluate the impact of the
effective mass on those observables. In order to obtain barrier heights, the 
computation of the energy landscape as a function of the quadrupole moment is required.
An example of such kind of calculations is shown in Fig \ref{fig:fission} using the paradigmatic case 
of $^{240}$Pu. For comparison, the results obtained with BCPM and with the Gogny D1S
functional \cite{D1S} are also plotted. The potential energy is given by the HFB one including the 
standard rotational energy correction 
$\langle \Delta \vec{J}^2 \rangle / (2 \mathcal{J}_\mathrm{Y})$ \cite{Ring.80}. 
The results for the three functionals
show a very similar behavior, with the position of
maxima and minima being almost the same in the three cases. It is remarkable to notice
the shoulder obtained with both BCPM and BCPM$^*$ at $Q_{20}=90 \mathrm{b}$ which is reminiscent
of a second isomeric well. The connection of this shoulder with the second isomeric well observed
in some U and Th isotopes deserves further investigation. The values obtained for
the two barrier heights are given in Table \ref{tbl:fission} along with the corresponding numbers
for $^{234}$U and $^{246}$Cm and compared with experimental data.

\begin{table}

\begin{tabular}{cccccccccc}
          & E$_A$ & E$_B$ & E$_I$  & E$_A$  & E$_B$ & E$_{I}$ & E$_A$ & E$_B$  & E$_{I}$  \\ 
          &\multicolumn{3}{c}{BCPM$^*$}& \multicolumn{3}{c}{BCPM}& \multicolumn{3}{c}{Exp}   \\ \hline \hline
$^{234}$U &   5   & 5.8   & 1.8     &   5.6 & 5.6   & 2.      &   4.8 & 5.5    & --       \\
$^{240}$Pu&   6.2 & 5.5   & 1.7     &   7.3 & 5.8   & 2.1     &   6   & 5.15   & 2.8      \\
$^{246}$Cm&   6.5 & 4.7   & 1.1     &   8   & 5.5   & 2.1     &   6   & 4.8    & --       \\ \hline
\end{tabular}
\caption{\label{tbl:fission} First (E$_A$) and second (E$_B$) fission barrier heights 
and the excitation energy of the
fission isomer (E$_I$) are given in MeV for three typical actinide nuclei. Results 
obtained with BCPM$^*$ and BCPM are given along with the experimental data from \cite{Ber15}.}
\end{table}

We observe that both the results obtained with BCPM$^*$ as well as the ones with BCPM are in a quite
good agreement with experimental data \cite{Ber15}, the ones obtained with BCPM$^*$ being slightly better. 
The results can not be taken as conclusive because triaxiality is not allowed to develop in the first barrier.
However, the agreement of the calculations with experimental data for the second barrier 
is very encouraging as,
in this case, triaxiality has proven to play a marginal role. Other observable
quantity relevant in fission studies is the excitation energy of the fission 
isomer E$_I$ which is also given in the Table
along with the only existing experimental datum for $^{240}$Pu. The theoretical predictions are lower than
the experimental value by around 25 \% to 40 \%, with the BCPM$^*$ result lower than the BCPM one and, therefore,  
slightly worse in terms of the comparison with the experiment.

In the three cases studied, reflection symmetry is broken for quadrupole
moments beyond the second fission barrier. The behavior and values of the octupole
moment for those configurations are very similar to the ones obtained
with BCPM and Gogny D1S, indicating that the good octupole properties of those
functionals are preserved in the present proposal. Work to analyze in more
detail deformation properties of BCPM$^*$ is in progress and will be reported
elsewhere. 

\subsection{Monopole and quadrupole giant resonance energies}

\begin{figure}
\includegraphics[angle=-90,width=\columnwidth]{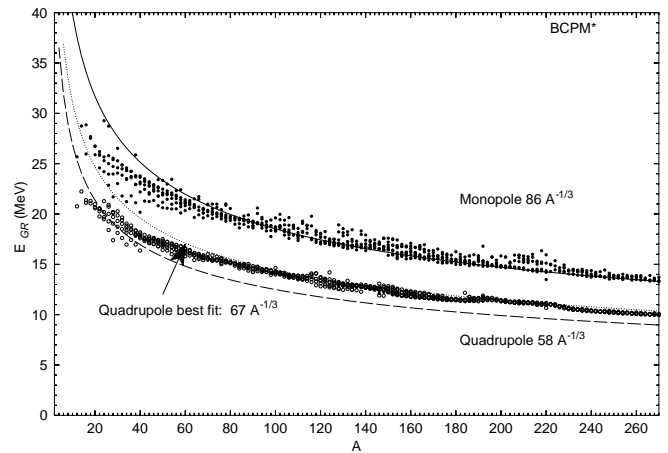}
\caption{Excitation energies of the monopole and quadrupole giant resonances as 
a function of mass number A obtained with the scaling approximation. The estimation $58 A^{-1/3}$ for the quadrupole and
$86 A^{-1/3}$ for the monopole \cite{Ring.80} are drawn to guide the eye. The dotted curve 
$67 A^{-1/3}$ corresponds to the quadrupole's best fit.\label{fig:EGR}}
\end{figure}

In our previous work we have discussed with some detail the excitation 
properties of the BCPM energy density functional. In particular, we 
analyzed the excitation energies of the scalar giant monopole and 
quadrupole resonances (GMR and GQR, respectively)
using sum rule techniques. We found that the BCPM predictions for 
the excitation energies of the GMR were in agreement with the results provided
by other mean field models, non-relativistic and relativistic, with 
a similar value of the incompressibility modulus $K$. However, it was found that the
experimental excitation energies of the GQR were systematically underestimated 
by about 1 MeV. The underlying reason for that was that in BCPM the effective
mass equals the bare one and it is known that the GQR excitation energies are
sensitive to the value of the effective mass.

We have repeated these calculations in the present work using the BCPM$^*$ 
energy density functional. In Table \ref{tab:EGR} we report the theoretical
estimates of the excitation energy of the GMR and GQR, computed with our
new functional, of a selected set of nuclei, for which the GMR excitation 
energy is experimentally known. 

Comparing with Table VII of \cite{BCPM}, one can see that the influence of
the effective mass on the excitation energy of the GMR is basically negligible,
while it is noticeable in the case of the GQR. This behavior can be understood 
in the scaling approach as follows. The scaled $m_3$ sum rules
for the GMR and the GQR (see Eqs. (A12) and (A17) of \cite{BCPM}) contain a 
kinetic energy contribution coming from the second derivative of the
scaled energy density respect to the scaling parameter $\lambda$.
According to the transformation of Eq. \ref{eq:kinem}, used in BCPM$^*$ to account for the
kinetic energy, it is easy to see that both contributions  $\tau$ and $\tau^{\infty}$ scale 
as $\lambda^2$ in the monopole case while in the quadrupole one only $\tau$, given by Eq. 
\ref{eq:quantumtau}, contributes as a consequence of volume conservation in the quadrupole 
oscillation. As far as
in BCPM$^*$ the effective mass $m^*$ is smaller than the bare mass $m$, the $m_3$ value and the excitation energy of 
the GQR $E_3(Q)$ will be larger than the predictions of the BCPM functional where $m^*=m$. 
Therefore, the agreement with the experimental values of the excitation energy of the GQR is better when
computed with the BCPM$^*$ functional, as it can be seen in Table II.

In Figure \ref{fig:EGR} we display the excitation energies of the monopole 
and quadrupole oscillations along the whole periodic table. Both follow a
$C\,A^{-1/3}$ law with coefficients $C_M=86$ and $C_Q=67$ MeV, respectively. 
These values are roughly in agreement with the empirical values given
in \cite{Ring.80} of 86 and 58 MeV for the monopole and quadrupole resonances. 


\begin{table}
\begin{center}
\caption{\label{tab:EGR} Theoretical $E_3$ and $E_1$ estimates (in MeV) of the average 
excitation energy of the GMR including pairing correlations. The $E_3$ 
estimate of the GQR, also including pairing, is also displayed. The 
experimental energy of the centroid and the corresponding error for the 
GMR and GQR are also given.}
\begin{tabular}{cccccc}
\hline \hline
 Nucleus & $E_3(M)$ & $E_1(M)$ & $E_3(Q)$ & Exp(M)           &     Exp(Q)        \\ \hline
$^{90}$Zr  &  19.10 &    18.31 &    14.65 & 17.81 $\pm$ 0.32 & 14.30 $\pm$ 0.40  \\
$^{144}$Sm &  16.45 &    15.66 &    12.59 & 15.40 $\pm$ 0.30 & 12.78 $\pm$ 0.30  \\
$^{208}$Pb &  14.53 &    13.89 &    11.18 & 13.96 $\pm$ 0.20 & 10.89 $\pm$ 0.30  \\
$^{112}$Sn &  17.73 &    16.96 &    13.64 & 16.1  $\pm$ 0.1  & 13.4  $\pm$ 0.1   \\
$^{114}$Sn &  17.62 &    16.85 &    13.56 & 15.9  $\pm$ 0.1  & 13.2  $\pm$ 0.1   \\
$^{116}$Sn &  17.50 &    16.81 &    13.48 & 15.8  $\pm$ 0.1  & 13.1  $\pm$ 0.1   \\
$^{118}$Sn &  17.39 &    16.70 &    13.41 & 15.6  $\pm$ 0.1  & 13.1  $\pm$ 0.1   \\
$^{120}$Sn &  17.28 &    16.59 &    13.34 & 15.4  $\pm$ 0.2  & 12.9  $\pm$ 0.1   \\
$^{122}$Sn &  17.18 &    16.39 &    13.28 & 15.0  $\pm$ 0.2  & 12.8  $\pm$ 0.1   \\
$^{124}$Sn &  17.08 &    16.28 &    13.22 & 14.8  $\pm$ 0.2  & 12.6  $\pm$ 0.1   \\
$^{106}$Cd &  18.01 &    17.16 &    13.87 & 16.50 $\pm$ 0.19 &                   \\ 
$^{110}$Cd &  17.74 &    16.89 &    13.68 & 16.09 $\pm$ 0.15 & 13.13 $\pm$ 0.66  \\    
$^{112}$Cd &  17.61 &    16.85 &    13.59 & 15.72 $\pm$ 0.10 &                   \\
$^{114}$Cd &  17.48 &    16.65 &    13.50 & 15.59 $\pm$ 0.20 &                   \\
$^{116}$Cd &  17.36 &    16.53 &    13.42 & 15.40 $\pm$ 0.12 & 12.50 $\pm$ 0.66  \\  
\hline \hline
\end{tabular}
\end{center}
\end{table}


\section{Conclusions}

In this work we propose  a variant of the BCPM energy density functional published 
in \cite{BCPM}, where the bare mass is replaced by a density dependent effective 
mass $m^{*}$. Though it may not be absolutely clear whether bare or effective mass 
is preferable as we argued in the Introduction, it is certainly true that for, 
e.g., giant resonances other than monopole and dipole ones an effective mass 
$m^* < m$ is favored. Again we used our strategy and deduced the effective 
mass from our microscopic G-matrix results and adjusted separately proton 
and neutron effective masses to our results of the Bruckner G-matrix 
using polynomials in the density. A linear interpolation between proton and 
neutron masses was fitted to the asymmetries prevailing in finite nuclei. 
It turns out that the difference of both masses is quite a bit on the lower 
side of what one generally finds in the literature. In finite nuclei the 
densities are then replaced by the local ones as obtained from the HFB calculation. 
Some parameters had to be readjusted, in first place this concerns the strength 
of the spin-orbit term, which now has values much closer to the usual values of 
Skyrme or Gogny functionals. Concerning the results, not surprisingly, the 
ones for the giant quadrupole resonance are now in significantly better 
agreement with experimental data. Fission barriers from BCPM$^{*}$ are 
slightly better than those with BCPM. On the other hand the excitation 
energy of the fission isomer is slightly worse for the only data point 
of $^{240}$Pu with BCPM$^*$ than BCPM. The rms value for the masses is 1.68  MeV
with BCPM$^*$ and 1.58 MeV with BCPM, the rms value for the radii is 0.024 fm instead of 0.027 fm.
All in all it can be said that BCPM$^*$ practically performs as well as BCPM 
for all quantities besides for the GQR where it yields sensitively better results.




\begin{acknowledgments}
Work supported in part by the Spanish MINECO grants Nos. FPA2012-34694, FIS2013-34479, and 
FIS2014-54672-P; by the Consolider Ingenio 2010
programs MULTIDARK CSD2009-00064 and CPAN CSD2007-00042, by the project
MDM-201-0369 of ICCUB from MINECO and by the grant 2014SGR-401 from the
Generalitat de Catalunya. 
\end{acknowledgments}

\end{document}